# Controlling the Interlayer Coupling of Twisted Bilayer Graphene

Lan Meng[1], Wei Yan[1], Zhao-Dong Chu[1], Yanfeng Zhang[2], Lei Feng[1], Rui-Fen Dou[1], Jia-Cai Nie[1], and Lin He[1,*]

[1] Department of Physics, Beijing Normal University, Beijing, 100875, People's Republic of China

[2] College of Engineering, Peking University, Beijing, 100871, People's Republic of China

The interlayer coupling of twisted bilayer graphene could markedly affect its electronic band structure. A current challenge required to overcome in experiment is how to precisely control the coupling and therefore tune the electronic properties of the bilayer graphene. Here, we describe a facile method to modulate the local interlayer coupling by adsorption of single molecule magnets onto the twisted bilayer graphene and report the characterization of its electronic band structure using scanning tunneling microscopy and spectroscopy. The low-energy Van Hove singularities (VHSs) and superlattice Dirac cones, induced by the interlayer coupling and graphene-on-graphene moiré respectively, are observed in the tunneling spectra. Our experiment demonstrates that the energy difference of the two VHSs, which reflects the magnitude of interlayer coupling, can be tuned by the local coverage density of adsorption.

## I. INTRODUCTION

The experimental studies of graphene were starting with the pioneering work of Novoselov *et al.* in 2004.[1] Since that time, there has been a continuously growing interest in both the fundamental physics and applications of this system.[2-7] Several years later, unabated new phenomena are still being uncovered in this feracious field.[4,7-9] Compared with single-layer graphene, bilayer graphene displays even more intriguing properties[10-19] including nematic phase transition[20] and low-energy Van Hove singularities (VHSs).[21-26] The interlayer electron hopping of bilayer graphene markedly affects its band structure and electronic properties.[13,14,16,21-29] Accordingly, controlling the interlayer coupling is becoming one of the central issue in the physics of bilayer graphene system.

Recently, Li *et al.* demonstrated that a rotation between stacked graphene layers can influence the interlayer hopping and lead to the appearance of two low-energy VHSs flanking the Dirac point.[21] This result not only opens intriguing prospects for VHS engineering of electronic phases but also reveals exciting opportunities for inducing and exploring new phases with desirable properties in graphene.[24-26] In a previous paper[24], we demonstrated that the magnitude of the interlayer coupling can be enhanced by adsorption of single molecule magnets (SMMs), $[Mn_{12}O_{12}(CH_3COO)_{16}(H_2O)_4]\cdot 2CH_3COOH\cdot H_2O$ ($Mn_{12}$-ac), on the twisted bilayer graphene.[30] The adsorption of $Mn_{12}$-ac induced local curvature variation of graphene, which is expected to shorten the distance between bilayer graphene and enhance the interlayer electron hopping locally. Two low-energy VHSs, induced by the enhanced interlayer coupling, are observed as two pronounced peaks in the tunneling spectra. Our experiment also indicated that the periodic AB stacked atoms (the A atom of layer 1 and the B atom of layer 2 that have the same horizontal positions) of twisted graphene bilayer with a finite interlayer hopping enhanced the intervalley scattering from one Dirac cone to the other.[24]

Bringing the potential of inducing and exploring correlated electronic phases in graphene bilayer to reality desires a facile method to precisely control the interlayer coupling and therefore tune the electronic structure and properties of the twisted graphene bilayer. In this paper, we show that it is possible to tune the interlayer coupling by the local coverage density of $Mn_{12}$-ac on the twisted bilayer graphene. By using scanning tunneling microscopy and spectroscopy (STM and STS), we find that adsorption of $Mn_{12}$-ac markedly enhances the interlayer coupling and alters the electronic band structure of the bilayer graphene. Our experiment demonstrates that the energy difference of the two VHSs, which reflects the magnitude of interlayer coupling, can be tuned by the local coverage density of $Mn_{12}$-ac. Additionally, possible evidence for the superlattice Dirac cones[31-33], induced by the graphene-on-graphene moiré of the twisted graphene bilayer, was observed in the tunneling spectra.

## II. EXPERIMENT

Epitaxial graphene bilayer was grown in ultrahigh vacuum by thermal Si sublimation on hydrogen-etched n-doped 6H-SiC(000-1) substrate with doping density of $10^{16}$ cm$^{-3}$, which was purchased from TanKeBlue Semiconductor Co. Ltd in China. The details of the synthesis were reported in our previous papers.[24,30] The as-grown sample was charectarized by micro-Raman spectroscopy, STM, and STS measurements. The thickness of the sample was mainly bilayer. Our experiments demonstrated that the bilayer graphene with a twist angle of ~ 4.5° (the period of moiré pattern is about 3.1 nm) behaves as single layer. The interlayer coupling strength is so weak that the electronic band structure is identical to that of single-layer graphene.[30] Later, we demonstrated that the adsorption of $Mn_{12}$-ac on the twisted bilayer graphene enhanced the interlayer coupling and strengthened the intervalley scattering.[24] In this paper, we studied carefully the electronic band structure of twisted bilayer graphene as a function of the local coverage density of $Mn_{12}$-ac. The $Mn_{12}$-ac powders were diluted in isopropanol solution with the help of supersonic cleaner.

Two drops of the isopropanol solution with the diluted $Mn_{12}$-ac powders were dripped on the epitaxial graphene. The sample was dried in a stream of dry nitrogen and was transferred into ultrahigh vacuum chamber for STM and STS studies. Next, the sample was transferred out of the STM chamber to increase the coverage density of the $Mn_{12}$-ac and the newly obtained sample was further studied by STM and STS measurements. The STM system was an ultrahigh vacuum four-probe scanning probe microscopy (SPM) from UNISOKU. All STM and STS measurements were performed at liquid-nitrogen temperature and the images were taken in a constant-current scanning mode. The STM tips were obtained by chemical etching from a wire of Pt(80%)Ir(20%) alloys. Lateral dimensions observed in the STM images were calibrated using a standard graphene lattice. The STS spectrum, i.e., the dI/dV-V curve, was carried out with a standard lock-in technique using a 957 Hz alternating current modulation of the bias voltage.

## III. RESULTS AND DISCUSSION

Figure 1(a) shows a STM image of the epitaxial twisted graphene bilayer modified with low coverage density of $Mn_{12}$-ac molecules. We can observe the clear moiré pattern with the period of 3.1 nm as pristine twisted graphene bilayer, which indicates that the chemisorption did not alter the twist angle of the bilayer graphene. The $Mn_{12}$-ac molecules disperse randomly on the surface of graphene. The height of most $Mn_{12}$-ac molecules is measured to be about 2.8 nm, which is slightly larger than the size of a single $Mn_{12}$-ac molecule ~ 2.0 nm.[24] This may be attributed to the influence of the protuberances of the moiré pattern and/or local curvature variation of graphene induced by the adsorption of $Mn_{12}$-ac molecules. The nominal coverage density of the $Mn_{12}$-ac molecules in Fig. 1(a) is estimated to be $1.32 \times 10^{-3}$ $nm^{-2}$. It was estimated by the total number of $Mn_{12}$-ac molecules, ~ 64, in an area of $4.86 \times 10^{4}$ $nm^{2}$ and the obtained coverage density should only be taken as a very rough estimate since the $Mn_{12}$-ac molecules are not uniformly distributed.

To study the effects of $Mn_{12}$-ac molecules on the electronic band structures of the twisted graphene bilayer, STS measurements were performed. The tunnelling spectrum gives direct access to the local density of states (LDOS) of the surface at the position of the STM tip. Fig. 1(b) shows four typical dI/dV-V curves obtained on the surface of the epitaxial bilayer graphene around the $Mn_{12}$-ac molecules. The curves from I to IV are STS spectra recorded at the positions labeled in Fig. 1(a). The obtained results indicate that the twisted bilayer graphene no longer shows single-layer behavior.[24,30] Instead, the spectra develop two sharp peaks flanking the Dirac points with energy separation ~ 0.50 eV. The intensity of the peaks varies at different positions, however, the energy difference of the two peaks is almost a constant, irrespective of the positions. The two sharp peaks in the spectra were attributed to the low-energy VHSs of the twisted graphene bilayer induced by interlayer coupling. The positions of the two VHSs flanking the Dirac points are not symmetrical about zero bias. This may arises from the slight departure between the Fermi level and the Dirac

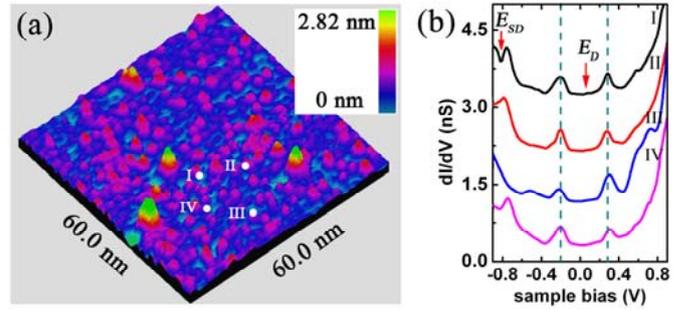

FIG. 1 (color online). (a) A typical STM image ($V_{sample}$ = 314 mV and I = 0.28 nA) of $Mn_{12}$-ac molecules adsorption on the surface of bilayer graphene, with the nominal coverage density of the $Mn_{12}$-ac molecules ~ $1.32 \times 10^{-3}$ $nm^{-2}$. The bright green dots are $Mn_{12}$-ac molecules with the height of about 2-3 nm. (b) Typical dI/dV-V curves obtained on the surface of the epitaxial bilayer graphene around the $Mn_{12}$-ac molecules. The curves from I to IV were measured at the positions labeled in the STM image. The red arrow near zero-bias of the spectra points to the Dirac point $E_D$ of twist bilayer graphene. The dip at about -0.8 V of the spectra is attributed to the superlattice Dirac points $E_{SD}$. Two peaks flanking the Dirac points, marked by green dot lines, are attribute to the low-energy VHSs induced by interlayer coupling.

point $E_D$ (pointed out by the red arrow in the dI/dV-V curves). For twisted graphene bilayer, the Dirac points of the two layers no longer coincide and there is a shift between the corresponding Dirac points ($K$ and $K_\theta$ for the first layer and the sub-layer respectively) in momentum space by an amount $\Delta K = 2K\sin(\theta/2)$, where $\theta$ is the twist angle and $K = 4\pi/3a$ with $a$ ~ 0.246 nm the lattice constant of the hexagonal lattice (the schematic Dirac cones of twisted graphene bilayer are shown in Fig. 2(a).). The displaced Dirac cones cross and, in the presence of interlayer hopping, two intersection of the saddle points (VHSs) are unavoidable along the two cones.[21-26] In the four-band model, four peaks including two low-energy VHSs can be obtained in the density of states (DOS) of the twisted bilayer graphene, as shown in Fig. 2(b). The

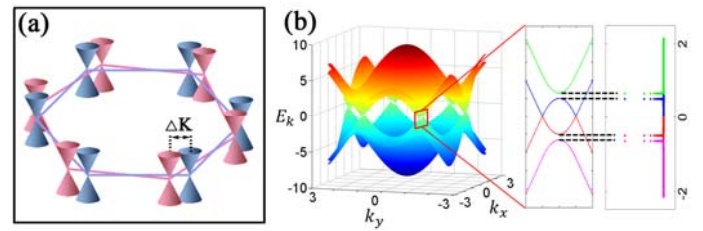

FIG. 2 (color online). (a) Schematic Dirac cones of twisted graphene bilayer. The separation of the Dirac cones $\Delta K$ in the two layers is attributed to the rotation between the graphene layers. (b) Left panel: electronic band structure of twisted bilayer graphene with a finite interlayer coupling calculated with the four-band model. Middle panel: a cut of electronic band structure along $K$ and $K_\theta$ in the low energy spectrum near the Dirac points. Right panel: peaks of density of states at the energy that $\triangledown E_k$ shows singularities.

energy difference of the two low-energy VHSs $\Delta E_{vhs}$ decreases with increasing the interlayer hopping, whereas, the opposite is the case for the energy difference between the two peaks in positive (or negative) bias (the two peaks include one of the VHSs and its nearest peak).

Besides the two VHSs, the tunneling curves in Fig. 2(b) show a dip in the LDOS (marked by an arrow) placed at about - 0.87 eV away from the graphene Dirac point. This dip is attributed to the superlattice Dirac cones induced by a weak periodic potential of the morié pattern.[31-33] Very recently, LeRoy and the co-authors demonstrated the emergence of the superlattice Dirac cones induced by the morié pattern between the graphene and the hexagonal boron nitride.[31] The presence of these superlattice Dirac points is manifested by two dips in the DOS at $E_{SD} \sim \pm \hbar v_F |G|/2$ (here $\hbar$ is the reduced Planck constant, $v_F$ the Fermi velocity of graphene, and $G$ the reciprocal superlattice vectors of the morié pattern), which are generally of asymmetric strength. Experimentally, the dip in the valence band is much deeper than that in the conduction band and the dip in the conduction band is usually not visible.[31] With considering the period of the graphene-on-graphene moiré $\sim$ 3.1 nm, $E_{SD}$ of the twisted graphene bilayer is estimated to be about $\pm$ 0.85 eV, which consists well with the experimental result. We will show more experimental evidences and analysis that support our understanding subsequently.

Figure 3(a) and (c) show two typical STM images of the sample with higher coverage density of $Mn_{12}$-ac molecules. The moiré pattern of the twisted bilayer graphene with the period of 3.1 nm can still be observed clearly. The nominal coverage density of $Mn_{12}$-ac in Fig. 3(a) and (c) is estimated to be $2.0 \times 10^{-3}$ $nm^{-2}$ and $3.0 \times 10^{-3}$ $nm^{-2}$ respectively. Fig. 3(b) and (d) show the corresponding dI/dV-V curves obtained on the surface of the epitaxial bilayer graphene around the $Mn_{12}$-ac molecules. Three characteristics can be easily observed from the tunneling curves: (I) the energy difference of the two low-energy VHSs $\Delta E_{vhs}$ decreases with increasing the coverage density of $Mn_{12}$-ac molecules; (II) the superlattice Dirac points are observed at $\sim$ -0.86 eV, irrespective of the coverage density of adsorption; (III) more peaks appear between the VHSs and the superlattice Dirac points $E_{SD}$ and the energy difference between the VHS and the nearest peak $\delta E$ increases with increasing the coverage density of $Mn_{12}$-ac molecules. Recently, it was predicted that strong interlayer coupling of twisted graphene could lead to many saddle points, and/or maxima, minima, and flat bands in the electronic band structure.[27-29] These Moiré bands also result in many singularities in the DOS, which may be the origin of the other peaks observed in the tunneling spectra.

Our experimental results and analysis are summarized in Fig. 4. Fig. 4(a) shows the $\Delta E_{vhs}$ and the $E_{SD}$ of the twisted bilayer graphene as a function of the coverage density of $Mn_{12}$-ac molecules. With increasing the coverage density of adsorption, the magnitude of $\Delta E_{vhs}$ decreases from about 0.50 eV to 0.23 eV. For twisted graphene bilayer with a small interlayer coupling, the energy difference of the VHSs can be roughly estimated by $\Delta E_{vhs} = \hbar v_F \Delta K - 2t_\perp$.[21] Here $v_F$ is the Fermi velocity of twisted bilayer graphene and $t_\perp$ is the interlayer hopping parameter. It indicates that the value of $\Delta E_{vhs}$ depends strongly on both the twist angle and the interlayer coupling. Very recently, several groups reported the twist angle dependent VHSs in twisted bilayer graphene.[21,25,26] In this paper, the twist angle is a constant ($\theta \sim 4.5°$). Therefore, the observed decrease of the magnitude of $\Delta E_{vhs}$ is attributed to the increase of interlayer coupling, which is induced by the chemisorptions of $Mn_{12}$-ac molecules. It means that the magnitude of interlayer coupling can be tuned by the local coverage density of $Mn_{12}$-ac. The curve in Fig. 4(b) shows the theoretical $\Delta E_{vhs}$ as a function of the interlayer coupling calculated by tight binding model. The expression $\Delta E_{vhs} = \hbar v_F \Delta K - 2t_\perp$ is valid only within very weak interlayer coupling. The interlayer hopping parameter is estimated as 0.27 eV, 0.42 eV, and 0.82 eV for nominal coverage density of $Mn_{12}$-ac molecules $1.32 \times 10^{-3}$ $nm^{-2}$, $2.0 \times 10^{-3}$ $nm^{-2}$, and $3.0 \times 10^{-3}$ $nm^{-2}$, respectively. With increasing the interlayer coupling, the magnitude of $\delta E$ also increase from 0.25 eV, as shown in

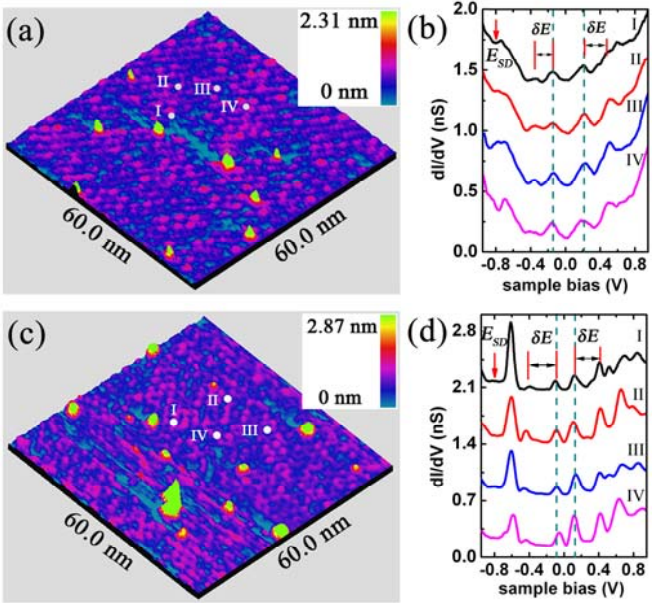

FIG. 3 (color online). (a,c) Two typical STM images with higher coverage density of $Mn_{12}$-ac molecules adsorption on the surface of the bilayer graphene (for panel (a) $V_{sample}$ = 302 mV and I = 0.39 nA; for panel (c) $V_{sample}$ = 393 mV and I = 0.42 nA). The bright green dots are $Mn_{12}$-ac molecules. The nominal coverage of the $Mn_{12}$-ac molecules is estimated to be (a) $2.0 \times 10^{-3}$ $nm^{-2}$ and (c) $3.0 \times 10^{-3}$ $nm^{-2}$. (b,d) Typical dI/dV-V curves obtained around the $Mn_{12}$-ac molecules on the surface of the epitaxial bilayer graphene. The curves from I to IV were measured at the positions labeled in the left STM images. All spectra shows two peaks flanking the Dirac points with energy separation $\Delta E_{vhs} \sim 0.36$ eV in panel (b), $\sim$ 0.23 eV in panel (d). The superlattice Dirac points are still observed at $\sim$ -0.8 eV, irrespective of the coverage density of adsorption. There are other peaks appear between the Van Hove singularities and the superlattice Dirac points $E_{SD}$. The energy difference between the Van Hove singularity and the nearest peak is denoted as $\delta E$.

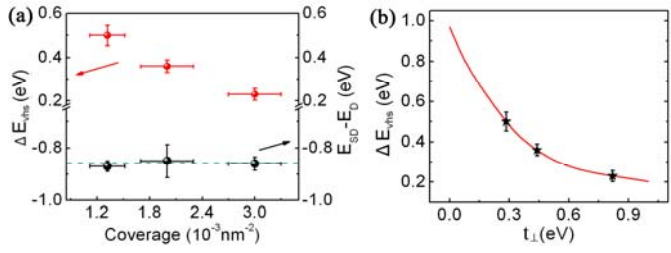

FIG. 4 (color online). (a) The $\Delta E_{vhs}$ (left Y-axis) and the positions of superlattice Dirac cone (right Y-axis) of the twist bilayer graphene as a function of the coverage density of $Mn_{12}$-ac molecules. The error bars in energy represent the minimum and maximum observed energies of $\Delta E_{vhs}$ and $E_{SD}$-$E_D$. The error bars in X-axis represent the uncertainty of the coverage density. (b) The solid curve is the theoretical energy separation of $\Delta E_{vhs}$ as a function of interlayer coupling calculated by tight binding model. The experimental $\Delta E_{vhs}$ is also plotted to estimated the interlayer coupling paremeter.

Fig. 3(b), to about 0.31 eV, as shown in Fig. 3(d).

On the basis of above results, it can be concluded that the chemisorption of $Mn_{12}$-ac molecules enhances the coupling between graphene bilayer, which may originate from the modulation of the distance between the bilayer. Further experiment should be carried out to explore the exact origin of the enhanced interlayer coupling induced by the chemisorption. The reported superlattice Dirac cones induced by graphene-on-graphene morié is also a very interesting topic. Further experiments to study the twisting-angle dependence of $E_{SD}$ could give unmistakable signature of the emergence of superlattice Dirac cones induced by graphene-on-graphene morié.

## IV. CONCLUSIONS

In summary, we report a facile method to modulate the interlayer coupling by absorption of SMMs onto the twisted bilayer graphene. The separation of the two VHSs flanking the Dirac point decreases with increasing the interlayer hopping (the nominal coverage density of adsorption). Meanwhile, evidence for the superlattice Dirac cones induced by the graphene-on-graphene moiré of the twisted graphene bilayer was observed in the tunneling spectra. The results reported here open opportunities for bringing the potential of inducing and exploring correlated electronic phases in graphene bilayer to reality.

## ACKNOWLEGEMENTS

This work was supported by the National Natural Science Foundation of China (Grant Nos. 10804010, 10974019, 11004010, 51172029 and 91121012), the Fundamental Research Funds for the Central Universities, and the Ministry of Science and Technology of China (Grants Nos. 2011CB921903).

*corresponding author helin@bnu.edu.cn .